**Possible superconductivity in Bismuth (111) bilayers. Its electronic and vibrational properties from first principles.**

David Hinojosa-Romero[1], Isaías Rodriguez[2], Alexander Valladares[2], Renela M. Valladares[2] and Ariel A. Valladares[1] *.
[1]Instituto de Investigaciones en Materiales, Universidad Nacional Autónoma de México, Apartado Postal 70-360, Ciudad Universitaria, CDMX, 04510, México.
[2]Facultad de Ciencias, Universidad Nacional Autónoma de México, Apartado Postal 70-542, Ciudad Universitaria, CDMX, 04510, México
* Corresponding Author: Ariel A. Valladares, valladar@unam.mx

**ABSTRACT**

Using a 72-atom supercell we report *ab initio* calculations of the electronic and vibrational densities of states for the bismuth (111) bilayers (bismuthene) with periodic boundary conditions and a vacuum of 5 Å, 10 Å and 20 Å. We find that the electronic density of states shows a metallic character at the Fermi level and that the vibrational density of states manifests the expected gap due to the layers. Our results indicate that a vacuum down to 5 Å does not affect the electronic and vibrational structures noticeably. A comparison of present results with those obtained for the Wyckoff structure is displayed. Assuming that the Cooper pairing potential is similar for all phases and structures of bismuth, an estimate of the superconducting transition temperature gives 2.61 K for the bismuth bilayers.

**INTRODUCTION**

Bulk bismuth is known to be a semimetal, a metal or a superconductor, with peculiar electronic and vibrational properties depending on whether it is crystalline or amorphous or depending on the pressure applied on it. At ambient pressure and temperature, it crystallizes in the Wyckoff structure, Bi-I, [1] with rhombohedral symmetry in which each atom has three equidistant nearest-neighbor atoms and three equidistant next-nearest neighbors slightly further away, resulting in a buckled 2D honeycomb bilayer lying perpendicular to the [111] crystallographic direction.

Bismuthene, or the bilayers (111) of bismuth, Bi (111), recently have been the subject of much interest and investigation as an example of non-carbon low-dimensional materials and the influence of this low dimensionality on their electronic and transport properties. It has been argued that in this layered form, bismuth has properties of topological insulators [2-4] which are bulk insulators with protected boundary states [5]. This state of matter appears when there is an inversion in the electronic bands of 2D materials caused by perturbations [6]. Among the perturbing agents the following are cited: strains in the systems, controlled quantum well width in the process of growth, doping [6, 7] and recently, for the Bi (111), the strength of the spin-orbit coupling [8, 9]. Although some questioning has appeared on the topological insulator nature of the bismuth bilayers, it seems nonetheless an interesting speculation that should be followed through to discern the veracity of such statements.

Here we introduce another proposal based on our results concerning the superconductivity in several bismuth crystalline phases both at ambient pressure and also at high pressures [10-12]; we claim that Bi (111) bilayers may display superconductivity at liquid helium temperatures.

When Bardeen, Cooper and Schrieffer (BCS) explained superconductivity back in 1957, two important concepts were introduced; the first one was the phonon-mediated electron Cooper pairing that occurs due to the vibrations in the material, giving rise to the transition to the superconducting state, and the second one was the coherent motion of the paired electrons that gives them the inertia to sustain electrical currents for a long time without dissipation [13]. Several variations of these ideas have appeared in the course of time and even different mechanisms that pretend to substitute the original ones.

Superconducting-like phenomena have been invoked in other realms of physics like nuclear and elementary particles where the pairing mechanism should be adequately chosen. It has also been ventured that in principle all materials may become superconductors if cooled down to low enough temperatures. To add to the universality of the superconducting phenomena, a recent paper has appeared where the idea of the Cooper pairing is extended to describe the behavior of a pair of photons [14]. In this paper, it is claimed that photon pairs exchange virtual vibrations leading to an attractive photon-photon interaction as for electrons in the BCS theory, although photons are bosons unlike the electrons.

We here show that summoning the corresponding electron and vibrational densities of states of a given structure, superconductivity may appear provided the Cooper attraction sets in. This elemental approach, if proven correct, would indicate that superconductivity in bismuth can be understood in a simple manner without invoking eccentric mechanisms.

Using this discourse, we recently demonstrated that the Wyckoff crystalline phase of Bi may become superconductive [10]. This phase had been investigated for several decades but no superconductivity was found. However, after our publication, an experimental group decide to investigate it at temperatures lower than the upper bound we calculated, $T_c \leq 1.3$ mK, and found that this phase is indeed superconductive at $T_c = 0.53$ mK [15]. In a recent publication, we have shown that a simple compression without allowing for changes in crystalline structures cannot predict the appearance of superconductivity of the phases of Bi under pressure [16]. It is clear that if modifications in the crystalline structures are permitted, then there will be corresponding changes in the electron density of states (eDoS), $N(E)$, and in the vibrational density of states (vDoS), $F(\omega)$, and these are the factors that we consider to explain superconductivity in the different phases and structures of bismuth.

**METHOD**

We now proceed to calculate and analyze the $N(E)$ and $F(\omega)$ for the Wyckoff and bismuthene structures looking for a justification to validate our surmise that the Bi (111) bilayers may become a superconductor. We shall estimate its superconducting transition temperature *a la* Mata-Pinzón *et al*. [10], streamlined in Ref. 10. In order to obtain the electronic and vibrational properties of the Wyckoff phase, we proceed as before [10] using a supercell with 240 atoms. For bismuthene, we use a slab model, Figure 1, which consists of a 72-atom supercell with 5 Å, 10 Å and 20 Å of vacuum spacing (separation d in Figure 1) between Bismuth (111) bilayers including periodic boundary conditions.

$N(E)$ and $F(\omega)$, are calculated using the DMol3 code which is part of the Dassault Systèmes BIOVIA Materials Studio suite [17]. To obtain the eDoS a single-point energy calculation was performed first using a double-numeric (dn) basis set and a fine mesh within the LDA-VWN approximation; an unrestricted spin-polarized calculation for the energy was carried out. Since bismuth is a heavy element with many electrons, the density-functional semi-core pseudo-potential (DSPP) approximation was used [18]. This pseudo-potential has been investigated by Delley where an all electron calculation is compared to the DSPP; the rms errors are essentially the same for both methods, 7.7 vs 7.5 [18]. Scalar relativistic corrections are incorporated in these pseudopotentials, essential for heavy atoms like Bi. Since DSPPs have been designed to generate accurate DMol3 calculations, it is expected that their use represents a good approximation; considerations of symmetry were left out. The parameters used in the calculations were the same for the respective *N(E)* and *F(ω)*, so meaningful comparisons can be made. For example, an energy convergence of $10^{-6}$ Ha was used throughout, the real space cutoff was set to 6.0 Å, and the integration grid was set to fine, so the calculations were carried out using a Monkhorst-Pack mesh of 3x3x1 in k-space. For the vibrational calculations, the finite-displacement approach was employed with a step size of 0.005 Å to obtain the Hessian using a finite-difference evaluation.

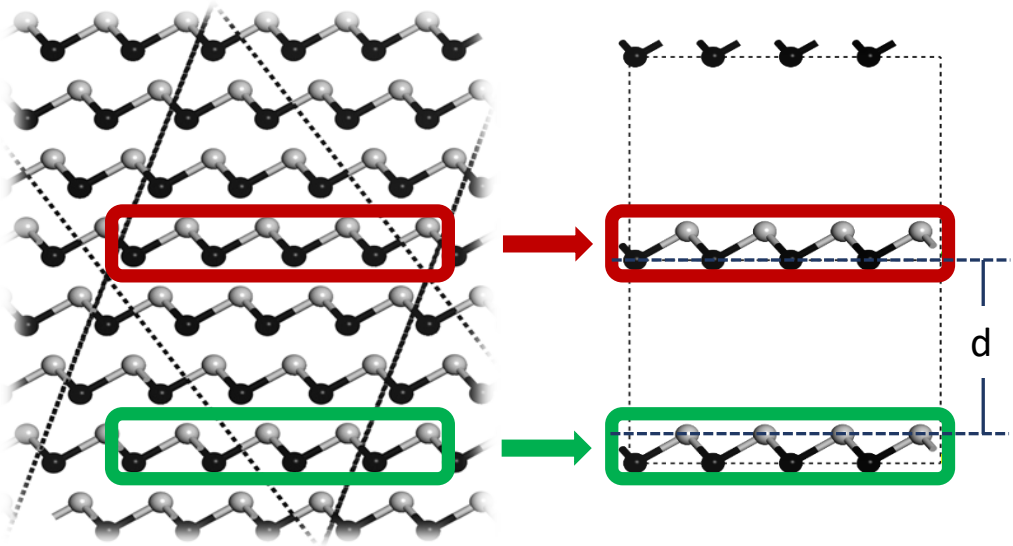

**Figure 1. Construction of the supercell for Bi (111).** Starting with the Wyckoff structure (left) we consider two identical slabs separated by a distance d (right). The bilayer is set to lie on the xy plane of the new supercell.

We now base our discussion on the BCS expression for the transition temperature:

$$T_c = 1.13\ \theta_D e^{\left(-\frac{1}{N(E_F)V_0}\right)}, \qquad (1)$$

where $\theta_D$ is the Debye temperature and represents the role played by the vibrational density of states, vDoS, typified by $F(\omega)$. $N(E_F)$ is the electron density of states, eDoS, at the Fermi level $E_F$, and $V$ is the Cooper pairing potential that binds pairs of electrons [13]. The dependence of $T_c$ on the parameters is represented in Fig. 2 for a specific arbitrary value of the pairing potential where the strong dependence of the transition temperature on the factor $N(E_F)$ can be observed.

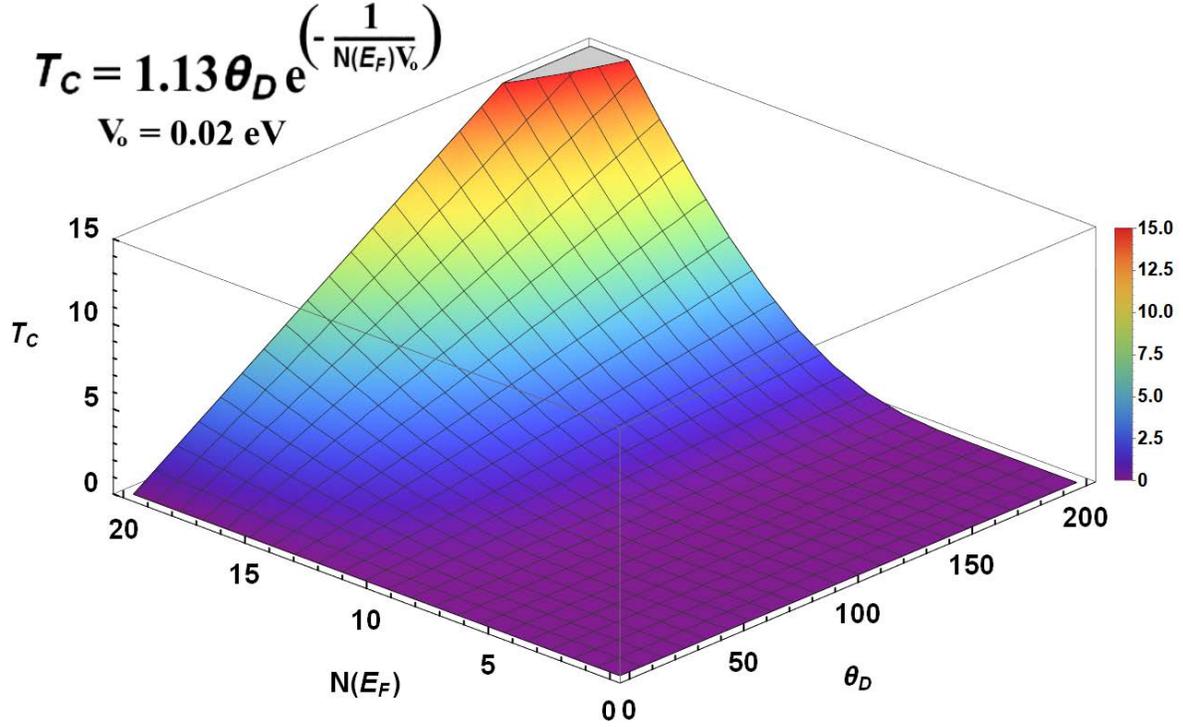

**Figure 2**. **Dependence of the superconducting transition temperature $T_c$ on the electronic $N(E_F)$ and vibrational $\theta_D$ parameters.** The pairing potential $V_o = 0.02$ eV was chosen arbitrarily.

This shows that under certain circumstances $N(E_F)$ can play a more important role than the vibrations, especially when different geometries of the same material are compared since in this case it can be assumed that the strength of the pairing potential would not be altered much by these changes. Although the Debye temperatures may not change drastically, the electronic properties could change radically going from being bulk to becoming a bismuthene, case at hand.

**RESULTS AND DISCUSSION**

For the eDoS the DMol3 analysis tools included in the Materials Studio suite were used, set to eV, and also an integration method with a smearing width of 0.2 eV. The number of points per eV was 100. The results for the densities of states are given per atom in Fig. 3, $N(E)/atom$.

For the vDoS the results were analyzed with the OriginPro software, the normal modes calculated were imported in THz. To obtain the vDoS a frequency count with a 0.11 THz bin width was used, and the resulting bins were smoothed with a two-point FFT filter. The three translational modes around 0 THz were removed. The results for the densities of states are given, per atom, in Fig. 4, $F(\omega)/atom$.

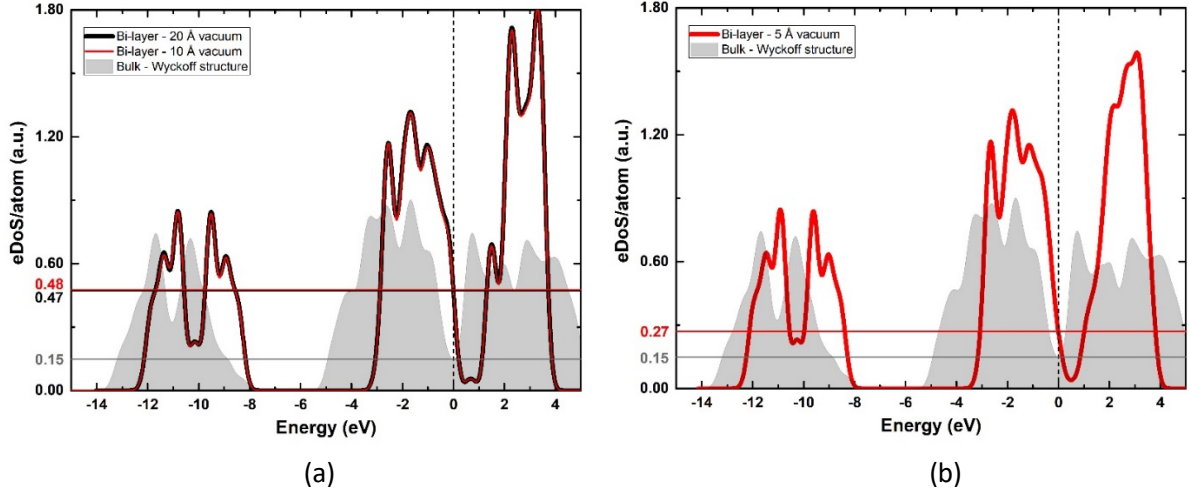

(a)                            (b)

**Figure 3**. **Comparison of the electron densities of states for the Wyckoff (gray plot) and Bi (111) structures (color lines).** (a) The eDoS for the layers $N(E)^L$ are practically the same for d= 10 Å and 20 Å. (b) When d= 5 Å $N(E)^L$ changes drastically at $E_F$. The vertical axis is $N(E)$/atom so that we can compare supercells of different sizes. $E_F$, the Fermi energy, is the reference value for the energy.

We now follow the reasoning of Ref. 9, streamlined in Ref 10. In Figure 3(a) the values for Bi (111) of $N(E)^L$ for d= 10 Å and 20 Å are compared to the Wyckoff $N(E)^W$ result. The $N(E)^L$ has a higher number of electron states, 0.48, at the Fermi level indicating a metallic character. $N(E)^W$, as expected, is lower in comparison since this phase is semimetallic, 0.15. The ratio of the densities of electron states at the Fermi level for these two phases is 3.20.

To incorporate the Debye temperatures as required by Eqn. (1) we need to calculate them from the vibrational spectra presented in Figures 4. For this we use an expression due to Grimvall [19],

$$\omega_D = \exp\left[1/3 + \frac{\int_0^{\omega_{max}} \ln(\omega) F(\omega) d\omega}{\int_0^{\omega_{max}} F(\omega) d\omega}\right], \qquad (2)$$

that we utilized in Refs. 9-11 with good results. $F(\omega)$ is the vDoS of the supercell under consideration and $\omega_{max}$ is the maximum frequency of the vibrational spectrum. Using Eqn. 2, the calculations for $\theta_D = \hbar\omega_D / k_B$ give: $\theta_D{}^L$ = 104.3 K and $\theta_D{}^W$ = 134.2 K, with $k_B$ the Boltzmann constant; the ratio being 0.78.

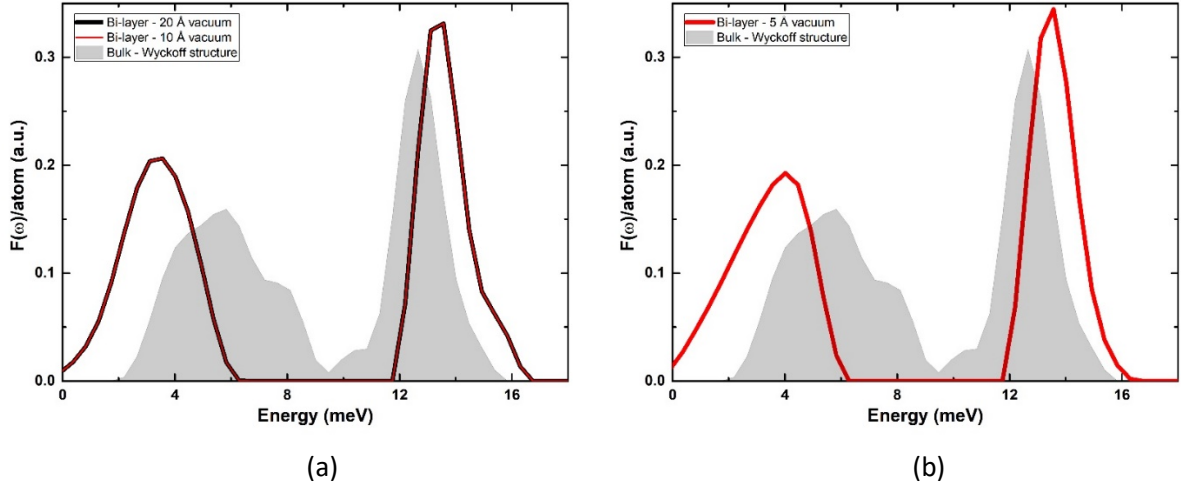

(a)                                            (b)

**Figure 4**. **Comparison of the vibrational densities of states for the Wyckoff (gray plot) and Bi (111) structures (color lines).** (a) The vDoS for the layers $F(\omega)^L$ are practically the same for d= 10 Å and 20 Å. (b) When d= 5 Å there are some minor changes in $F(\omega)^L$. The vertical axis is $F(\omega)$/atom so that we can compare supercells of different sizes.

We should mention that to obtain these results we removed the translational modes ($\omega \approx 0$) that are more preponderant the smaller the number of atoms in a supercell. That is why our present numbers are somewhat larger than those reported in Ref. 9 for the Wyckoff phase, 129 K. The experimental values for $\theta_D$ reported by DeSorbo [20] for crystalline bismuth at ambient pressure varies from 140 K at high temperatures to 120 K at low temperatures. Our calculations indicate that $\theta_D$ for the crystal at ambient pressure lies between the experimental values so we trust the results of 134.2 K for Wyckoff and 104.3 K for Bi (111).

Now we may compare the possible $T_c$ of the Bi (111) structure with the $T_c^W$ of the Wyckoff phase that was recently discovered to be superconductive. To be consistent, we do it with the superconductivity we predicted for the Wyckoff phase in the BCS approach. Suppose that superconductivity is possible for Bi (111), with a superconducting transition temperature $T_c^L$, and assume that the Cooper pair potential $V$ is essentially the same for these two structures of the same material. Then the transition temperatures are:

$$T_c^L = 1.13\, \theta_D^L e^{\left(-\frac{1}{N(E_F)^L V_0}\right)}$$

$$T_c^W = 1.13\theta_D^W e^{\left(-\frac{1}{N(E_F)^W V_0}\right)}.$$

If we assume that

$$N(E_F)^L = \alpha\, N(E_F)^W \quad \text{and} \quad \theta_D^L = \beta\, \theta_D^W,$$

we can rewrite the expression for $T_c^L$ as:

$$T_c^L = \{T_c^W\}^{1/\alpha} \beta \left(1.13\theta_D^W\right)^{(\alpha-1)/\alpha},$$

and substituting the values for $α = 3.20$ and $β = 0.78$ for our case, we obtain

$$T_c{}^L = 2.61 \text{ K}$$

## CONCLUSIONS

The bilayers of bismuth, or bismuthene, are very interesting because they form the basic structure of bulk bismuth, and they may display interesting properties when isolated. That is why we have investigated their electronic and vibrational properties as a function of the interlayer separation. We find that for interlayer distances higher than or equal to d = 5 Å the properties of the layers are practically identical, which means that the interaction between them is practically negligible. It makes sense that as the layers approach one another, the properties will resemble those of bulk bismuth although to obtain the Wyckoff phase a translation has to occur when interlayer distances approach the bulk ones.

Assuming that the Cooper pairing potential remains essentially unchanged in going from the bulk structure to the layer structure we predict, based on the changes mainly in the eDoS of these structures, that the layer will become superconductive with a $T_c{}^L = 2.61$ K.

## ACKNOWLEDGMENTS

D.H.R. and I.R. acknowledge CONACyT for supporting their graduate studies. A.V. R.M.V. and A.A.V. thank DGAPA-UNAM for continued financial support to carry out research projects IN110914 and IN104617. M.T. Vázquez and O. Jiménez provided the information requested. Simulations were partially carried out in the Computing Center of DGTIC-UNAM.